\documentclass[physics,article,submit,pdftex,moreauthors]{Definitions/mdpi} 
\renewcommand{\linenumbers}{}
\usepackage{subcaption}
\usepackage{graphicx}

\firstpage{1} 
\pubvolume{1}
\issuenum{1}
\articlenumber{0}
\pubyear{2023}
\copyrightyear{2023}
\datereceived{ } 
\daterevised{ } 
\dateaccepted{ } 
\datepublished{ } 
\hreflink{https://doi.org/} 

\Title{3D - 2D crossover and phase shift of beats of quantum oscillations of interlayer magnetoresistance in quasi-2D metals}

\TitleCitation{3D - 2D crossover and phase shift of beats of quantum oscillations of interlayer magnetoresistance in quasi-2D metals}

\Author{Taras Igorevich Mogilyuk $^{1}$\orcidA{}, Pavel Dmitrievich Grigoriev$^{2,\,3,\,4}$*\orcidB{}, Vladislav Dmitrievich Kochev$^{3}$\orcidD{}, Ivan Sergeevich Volokhov$^{1,\,3,\,6}$\orcidE{}, 
Ilya Yakovlevich Polishchuk$^{1,\,5}$\orcidC{}}

\AuthorNames{Taras Mogilyuk, Pavel Grigoriev, Vladislav Kochev, Ivan Volokhov and Ilya Polishchuk}

\AuthorCitation{Mogilyuk, T.I.; Grigoriev, P.D.; Kochev, V.D.; Volokhov I.S.; Polishchuk, I.Ya.}

\address{%
	$^{1}$ \quad Department of Condensed Matter, NRC Kurchatov Institute, Kurchatov Sq. 1, Moscow,
	Moscow, Russia 123182; 5taras@mail.ru; volohov000@mail.ru\\
	$^{2}$ \quad L. D. Landau Institute for Theoretical Physics of the Russian Academy of Sciences,
	Prospekt Ak. Semenova -- 1a, Chernogolovka, Moscow region, Russia 142432 \\
	$^{3}$ \quad Department of Theoretical Physics and Quantum Technology, National University of Science and Technology ”MISiS”,
	Leninskiy prospekt -- 4, Moscow, Russia 119049; vd.kochev@gmail.com\\
	$^{4}$\quad Kotelnikov Institute of Radioengineering and Electronics of RAS, 125009 Moscow, Russia\\
	$^{5}$\quad Theoretical Physics Department, Moscow Institute For Physics and Technology, Institutskii per. 9, Dolgoprudnii, Moscow Region, Russia 141700; iyppolishchuk@gmail.com\\
	$^{6}$ \quad Lebedev Physical Institute of the Russian Academy of Sciences, 119991, Moscow, Leninsky Avenue, 53}

\corres{Correspondence: grigorev@itp.ac.ru}

\abstract{Magnetic quantum oscillations (MQO) are traditionally applied to investigate the electronic structure of metals. In layered quasi-two-dimensional (Q2D) materials the MQO have several qualitative features giving additional useful information, provided their theoretical description is developed. Within the framework of the Kubo formula and the self-consistent Born approximation, we reconsider the phase of beats in the amplitude of Shubnikov oscillations of interlayer conductivity in Q2D metals. We show that the phase shift of beats of the Shubnikov (conductivity) oscillations relative to the de Haas - van Alphen (magnetization) oscillations is larger than expected previously and, under certain conditions, can reach the value of $\pi/2$, as observed experimentally. We  explain the phase inversion of MQO during the 3D - 2D crossover and predict the decrease of relative MQO amplitude of interlayer magnetoresistance in a strong magnetic field, larger than the beat frequency.}


\begin{document}


\section{Introduction}
\begin{figure}[ht ]
	\centering
	\begin{subfigure}[t]{0.45\textwidth}
		\centering
		\includegraphics[scale=1.2,height=2in]{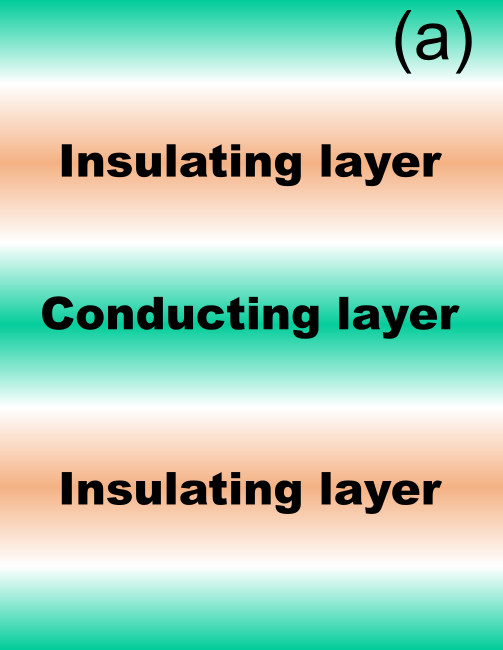}	
	\end{subfigure}
	\begin{subfigure}[t]{0.45\textwidth}
		\centering
		\includegraphics[scale=1.2,height=2in]{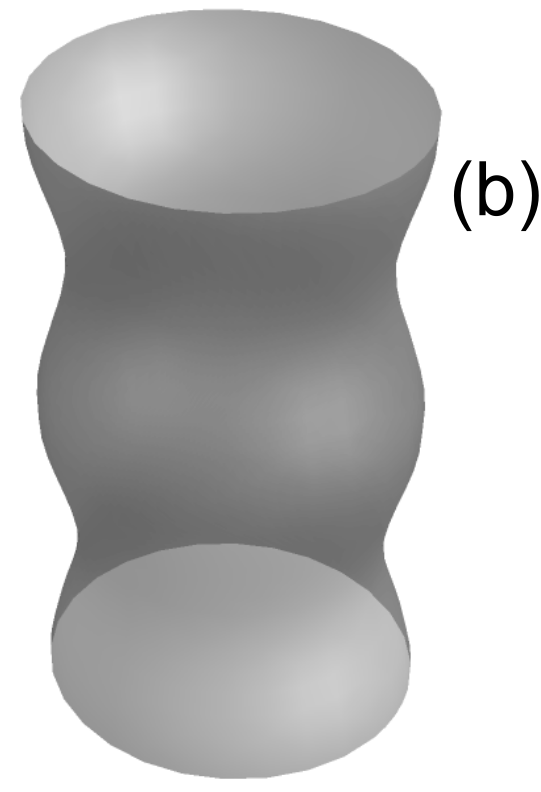}
	\end{subfigure}
	\caption{ Schematic representation of a layered Q2D metal (a) and its Fermi surface (b).
		\label{fig:0}}
\end{figure}

Layered quasi-two-dimensional (Q2D) metals (see Fig. \ref{fig:0} for illustration) represent a wide class of materials, intermediate between the various two-dimensional (2D) electron systems and usual three-dimensional (3D) compounds. The Q2D materials attract enormous research interest due to the variety of new electronic phenomena and diverse potential applications. They include high-temperature superconductors, organic metals, van der Waals crystals, artificial heterostructures, etc.
An experimental study of their electronic structure, in addition to ab initio calculations, is crucial for the understanding and utilizing their properties. Although the angle-resolved photoemission spectroscopy (ARPES) provides visual data on electronic bands and Fermi surface (FS), more traditional tools, such as magnetic quantum oscillations (MQO) \cite{Shoenberg1984Jan}, usually have higher accuracy and availability. 

The MQO originate from the Landau-level (LL) quantization of electron spectrum in a magnetic field $\boldsymbol{B}$ and appear in all thermodynamic and transport quantities as their periodic dependence on $1/B$. The standard MQO theory, based on the Lifshitz-Kosevich (LK) formula \cite{Shoenberg1984Jan}, is often applied even in Q2D metals and gives the FS extremal cross-section areas $S_i= 2\pi e\hbar F_{i}/c$ from the fundamental MQO frequencies $F_i$. Such data, collected at various directions of the magnetic field $\boldsymbol{B}$, helped to determine the FS of most known metals. Less accurate, the LK theory also describes the damping of MQO amplitude by temperature and disorder as a function of magnetic field strength $B$, which allows the estimate of electron effective mass and mean free time for each band \cite{Shoenberg1984Jan}. 

The strong anisotropy of layered Q2D metals introduces special features to MQO. 
The Q2D electron dispersion, obtained in the tight-binding approximation for 
the interlayer $z$ direction, is given by 
\begin{equation}
	\epsilon _{3D}(\boldsymbol{k})=\epsilon _{\parallel }\left( \boldsymbol{k}%
	_{\parallel }\right) -2t_{z}\cos (k_{z}d),  \label{e3Dg}
\end{equation}%
where  $\epsilon
_{\parallel }\left( \boldsymbol{k}_{\parallel }\right) $ is the intralayer electron dispersion, 
$t_{z}$ is the interlayer transfer integral of electrons, and $d$ is the lattice constant along the $z$-axis. 
Often in Q2D metals the Fermi energy $\mu \gg t_z$, and the FS is a warped cylinder with two close cross-section areas corresponding to the ''neck'' and ''belly'' of the FS (see Fig. \ref{fig:0}b). According to the LK theory, at $\boldsymbol{B}||z$ the MQO have two close fundamental frequencies $F_1=F-\Delta F/2$ and $F_2=F+\Delta F/2$. As the MQO amplitudes with these two frequencies are also close, the resulting MQO at frequency $F$ have amplitude modulations with frequency $\Delta F$, called beats. The beat frequency $\Delta F$ is proportional to the interlayer transfer integral: $\Delta F /F\approx 4t_z /\mu$.
It non-monotonously depends on the direction of magnetic field, which follows the angular magnetoresistance oscillations (AMRO) \cite{Yagi1990Sep,Moses1999Sep,Kartsovnik2004Nov,Wosnitza2013Fermi} and the can be used to determine the in-plane Fermi momentum \cite{Kartsovnik2004Nov,Wosnitza2013Fermi} and even high harmonics of FS warping \cite{Bergemann2000,Grigoriev2010}. 

Even when the MQO are weak, the interlayer electron transport in Q2D metals exhibits several unique properties, such as the aforementioned AMRO \cite{Yagi1990Sep,Moses1999Sep,Kartsovnik2004Nov,Wosnitza2013Fermi} and longitudinal interlayer magnetoresistance \cite{Grigoriev2011Jun,Grigoriev2012Oct,Grigoriev2013Aug}.   
There are even more new features of MQO in Q2D metals which are not described within the standard L-K theory. The interplay between the angular and quantum magnetoresistance oscillations is nontrivial \cite{Grigoriev2014Sep,Grigoriev2017May} and leads to the angular modulations of MQO amplitude \cite{Grigoriev2017May}. The so-called difference or slow oscillations (SO) of magnetoresistance \cite{Kartsovnik2002Aug,Grigoriev2003Apr,Grigoriev2017Oct,Mogilyuk2018Jul} at frequency $2\Delta F$ appear and survive at much higher temperature than the usual MQO. The SO help to determine the type of disorder and give other useful information, such as the interlayer transfer integral. Another feature beyond the L-K theory is the phase shift of beats between the MQO of thermodynamic and transport quantities \cite{Grigoriev2002Jan,Grigoriev2003Apr}. This phase shift increases with magnetic field strength $B$, as observed in various Q2D metals \cite{Grigoriev2002Jan,Schiller2000Jul,Weiss1999Dec} and substantiated theoretically \cite{Grigoriev2002Jan,Grigoriev2003Apr}. The experimentally observed phase shift of beats $\Delta\phi_{exp}$, usually, considerably exceeds the theoretical prediction $\Delta\phi_{theor}$ (e.g., see Figs. 3 of Ref. \cite{Grigoriev2003Apr} or Fig. 6 of Ref. \cite{Schiller2000Jul}). In the present paper we explain this inconsistency and develop a more accurate theory of this effect.

The beatings of MQO amplitude due to the interlayer electron hopping were observed in many other layered materials \cite{Kartsovnik2004Nov,Kang2010,Wosnitza2013Fermi,Danica2011,Audouard2015,Arnold2020,Oberbauer2023}. Note that the phase shift of beats is observed not only between the SdH and dHvA effect, but also between the oscillations of other quantities, e.g. the interlayer magnetoresistance and thermopower \cite{Danica2011}, or magnetoresistance and specific heat \cite{Wang2023}. The MQO of magnetization and of interlayer conductivity were also measured simultaneously in the Dirac semimetals BaGa$_2$ \cite{Xu2020}. There are many other combined SdH and dHvA measurements in layered metals where the MQO beats are observed \cite{Helm2021,Zeng2022,Luo2024,Daschner2024probing}. Similar beating effects may also appear in artificial heterostructures and multilayer topological insulators \cite{Leeb2021,Alisultanov2023}, as well as in layered Weyl semimetals \cite{Wang2023}. 

Another interesting feature of the 3D - 2D crossover, driven by the parameter $\lambda =4\Delta F/B_z $, is the phase inversion of MOQ of conductivity as compared to those of the electronic density of states (DoS) at the Fermi level and, hence, of all thermodynamic quantities. At low magnetic field the MQO of conductivity and of DoS have opposite phases, because the former is proportional to the electron mean free time which is inversely proportional to the DoS \cite{Abrikosov1988Fundamentals,Ziman1972Principles}, according to the golden Fermi rule or the Born approximation. However, in a high magnetic field $B_z \gg \Delta F$ in Q2D metals, when the Landau levels (LLs) become separated and the DoS between the LLs is zero, the MQO phase of conductivity and DoS coincide. The latter follows both from the direct observations \cite{Wosnitza2001,Grigoriev2012Oct,Laramee2023} and calculations \cite{Champel2002Nov,Grigoriev2011Jun,Grigoriev2012Oct,Grigoriev2013Aug,Mogilyuk2018Jul}, being supported by the simple qualitative argument that if the DoS at the Fermi level is zero the conductivity must also be zero at low temperature. However, in spite of a general understanding, the quantitative description of this phase inversion is absent. Below we show how this phase inversion during the 3D - 2D crossover is described by analytical formulas. The understanding of this MQO phase inversion is helpful for experimental studies of the Berry phase using MQO measurements \cite{Datta2019Oct,Xu2019,Zhao2022Dec,Nie2023Oct}. 

\section{Available experimental observations and their description}

	The MQO have been extensively investigated in various Q2D metals, including the cuprate \cite{Sebastian2012,VIGNOLLE2013,Helm2009,Breznay2019} and iron-based \cite{Carrington2011,COLDEA2013} high-Tc superconductors, organic metals \cite{Kartsovnik2004Nov,Wosnitza2013Fermi}, van-der-Walls layered crystals, artificial heterostructures, and many other materials. The two most common MQO experiments are the measurements of magnetization and conductivity oscillations, i.e. the de Haas - van Alphen (dHvA) and the Shubnikov - de Haas (SdH) effect. However, the observation of the phase shift between their beats is less common because it requires the measurements of both these effects on the same sample. 
	
	In Ref. \cite{Schiller2000Jul} the Fermi surface of a layered organic metal (BEDT-TTF)$_4$[Ni(dto)$_2$] was studied by measuring the quantum oscillations of both magnetization and interlayer conductivity. The authors have shown that the standard Lifshitz-Kosevich theory (LK) \cite{Shoenberg1984Jan} well describes the results on magnetization oscillations. The detected positions of the beat nodes of the de Haas - van Alphen oscillation amplitude are well fitted by the formula 
\begin{gather}
	B_{node}=\frac{4}{4n-1}\Delta F,\,n=1,\,2,\,3\ldots
	\label{eq:Bnote-dHvA}
\end{gather}
where $\Delta F$ is the beat frequency giving the FS warping.
However the observed position of the beat nodes of the Shubnikov oscillation amplitude in Ref. \cite{Schiller2000Jul} are strongly shifted and well fitted by another formula
\begin{gather}
	B_{node}=\frac{4}{4n+1}\Delta F,\,n=0,\,1,\,2\ldots
	\label{eq:Bnote-SdH}
\end{gather}
A similar difference in the positions of beat nodes of de Haas van Alphen oscillations and Shubnikov oscillations was also observed in
$\kappa-\textrm{(BEDT-TTF)}_2\textrm{Cu[N(CN)}_2]\textrm{Br}$
\cite{Weiss1999Dec}. The LK theory does not explain this clearly observed phase shift of beat nodes about $\pi/2$ between the oscillations of magnetization and interlayer conductivity. 

The 3D $\to $ 2D crossover in MQO behavior happens \cite{Grigoriev2012Oct} when $B\sim \Delta F$, i.e. when the LL separation $\hbar\omega_c$ becomes comparable to the interlayer bandwidth $4t_z$, where $\omega_c=eB/(m^{\ast}c)$ is the cyclotron frequency, $e$ -- the elementary charge, $c$ -- the speed of light, $m^{\ast}$ -- the effective electron mass, $\hbar$ --  the Planck constant. The phase shift between the MQO beats of dHvA and SdH effects is considerable when $\hbar\omega_c\gtrsim 4t_z$, i.e. during this 3D $\to $ 2D crossover. 

The expressions for interlayer conductivity derived \cite{Grigoriev2002Jan,Kartsovnik2002Aug} in the tau approximation strongly underestimate the observed phase shift $\phi_b$ between the beats of magnetization and interlayer conductivity, giving
\begin{equation}
	\phi_b=\arctan [\hbar\omega_c/(2\pi t_z)], 
		\label{PhSh1}
\end{equation}
   This formula always predicts $\phi_b<\pi /2$, which does not explain the experimental data from \cite{Schiller2000Jul}. 	
	Note that the phase shift of beats observed in Ref. \cite{Grigoriev2002Jan} is also $\sim 2.4$ times larger than the prediction of this formula. Indeed, substituting the known parameters in the quasi-two-dimensional organic metal $\beta $-(BEDT-TTF)$_2$IBr$_2$, namely, the cyclotron mass $m^{\ast} =4.2m_e$, as obtained from the temperature dependence of MQO amplitude, and the interlayer bandwidth $4t_z = 1.15$ meV, as extracted  from the ratio between the beating and fundamental MQO frequencies, to Eq. (\ref{PhSh1}) one obtains the $2.4$ times smaller slope of the linear dependence of $\tan \phi_b $ on $B_z$ than the experimental data in Fig. 4 of Ref. \cite{Grigoriev2002Jan}.
	
	The calculations of interlayer conductivity using the Kubo formula predict a larger phase shift of beats, given by \cite{Grigoriev2003Apr} 
\begin{equation}
	\tan \phi_{b}=\frac{\hbar \omega _{c}}{2\pi t_z}
	\left( 1+\frac{2\pi ^{2}k_{B}T_{D}}{\hbar \omega _{c}}\right) 
	=\frac{\hbar \omega _{c}}{2\pi t_z} \left( 1+\frac{\pi }{ \omega _{c}\tau }\right) 	\equiv \frac{\hbar \omega _{c}}{2\pi t_z} \left( 1+\gamma_0 \right), 
	\label{PhSh2}
\end{equation}
where $T_D\equiv\hbar/(2\pi k_B \tau)$ is the so-called Dingle temperature, $\tau $ is the electron mean free time, and $\gamma_0=\pi / (\omega _{c}\tau )$ is the dimensionless parameter. 
These quantities enter the Dingle factor $R_{D0}=\exp (-\gamma_0)=\exp [-2\pi ^{2}k_{B}T_{D}/(\hbar \omega _{c})]$, describing the damping of MQO by crystal disorder (see Eqs. (\ref{eq:M0})-(\ref{eq:M}) below). However, even this enhanced phase shift of beats, given by Eq. (\ref{PhSh2}), is still insufficient to describe quantitatively the effect observed in Refs. \cite{Schiller2000Jul,Grigoriev2002Jan}.

Note that the electron mean free time $\tau$ measured from the MQO is, usually, shorter than the transport mean free time $\tau_{tr}$ and than the mean-free time $\tau^{\ast}$ arising from short-range disorder only and measured from the damping of slow magnetoresistance oscillations \cite{Kartsovnik2002Aug}. This difference, $\tau <\tau^{\ast} \approx \tau_{tr}$, appears because the MQO are damped not only by short-range disorder but also by long-range sample inhomogeneities which smear the Fermi energy similar to the temperature effect \cite{Kartsovnik2002Aug,Grigoriev2003Apr,Grigoriev2012Oct,Grigoriev2017Oct,Mogilyuk2018Jul}. The opposite case when the transport mean free time $\tau_{tr}<\tau \lesssim \tau^{\ast} $ is also possible in heterogeneous conductors, where there are rare but strong inhomogeneities as, e.g., domain walls or linear crystal defects. Then inside each domain of the size larger than the Larmor radius the material is a clean metal with a large $\tau \sim \tau^{\ast} $, but the electronic transport across the sample is difficult because of its heterogeneity, which corresponds to a short $\tau_{tr}$. As follows from the calculations in Refs. \cite{Grigoriev2003Apr,Mogilyuk2018Jul}, it is $\tau^{\ast}$ rather than $\tau$ enters the Eq. (\ref{PhSh2}).  

\section{Materials and Methods}

The materials to which our study is applicable are the strongly anisotropic layered quasi-2D metals, where the interlayer transfer integral $t_z$ is of the same order of magnitude as the LL separation $\hbar\omega_c$. For the comparison of our formulas with experiment we take the data from Refs. \cite{Schiller2000Jul,Grigoriev2002Jan} obtained on the organic metals (BEDT-TTF)$_4$[Ni(dto)$_2$] and $\beta $-(BEDT-TTF)$_2$IBr$_2$. 

Below we take the magnetic field $\boldsymbol{B}$ perpendicular to conducting layers, so that $B=B_z$. We will use the expressions for the MQO of magnetization and interlayer conductivity from Refs. \cite{Champel2001Jan,Grigoriev2001Jun,Grigoriev2003Apr,Mogilyuk2018Jul} obtained in the self-consistent Born approximation. In the leading order of the expansion in powers of the Dingle factor $R_{D0}$ at temperature $T$, the oscillating parts of the magnetization is equal to (see Eq. (33) of Ref. \cite{Champel2001Jan} or Eq. (6) of Ref. \cite{Grigoriev2001Jun})
\begin{gather}
	\widetilde{{M}}\approx \frac{e\mu}{2\pi^2 \hbar c d}R_{D0}R_{T} \left[ J_{0}(\lambda )\sin\left(\overline{\alpha}\right)+\frac{\lambda }{%
		\overline{\alpha} }J_{1}(\lambda )\cos\left(\overline{\alpha}\right)\right] , \label{eq:M0} 
\end{gather}
and of interlayer conductivity is (see Eq. (18) of Ref. \cite{Grigoriev2003Apr} or Eq. (64) of Ref.  \cite{Mogilyuk2018Jul})
\begin{equation}
	\sigma_{zz}^{QO}\approx2\overline{\sigma}_{zz}^{(0)}\cos\left(\overline{\alpha}%
	\right)R_{D0}R_T  R_{z\sigma },\ \  R_{z\sigma }= J_{0}\left(\lambda\right)-\frac{2}{\lambda}%
	\left(1+\gamma_{0}\right)J_{1}\left(\lambda\right) ,  \label{eq:szzqo}
\end{equation}
where $R_{D0}=\exp(-\gamma_0) $ is the Dingle factor, $\gamma_0\equiv2\pi\Gamma_0/(\hbar\omega_c) = \pi /(\omega_c\tau )$, $\lambda  = 4\Delta F/B_z = 4\pi t_z/(\hbar\omega_c) $,  $\overline{\alpha}=2\pi \mu/(\hbar\omega_c)$,  $J_0$, $J_1$ -- the Bessel functions of zero and first order, $\overline{\sigma}_{zz}^{(0)}$ -- the interlayer conductivity without magnetic field,
$R_T=2\pi^2 k_B T/(\hbar\omega_c)/\sinh(2\pi^2 k_B T/(\hbar\omega_c))$ is the temperature damping factor of MQO.

At strong Q2D anisotropy, when $2t_z/\mu =\lambda /\overline{\alpha} \ll 1$ and the beats of MQO are the most pronounced, the last term in the square brackets of Eq. (\ref{eq:M0}) can be omitted, and the oscillating part of magnetization simplifies to
\begin{gather}
	\widetilde{{M}}\approx  \frac{e\mu}{2\pi^2 \hbar c d}\sin\left(\overline{\alpha} \right) R_{D0}R_{T} R_{M} ,\ \ 
	R_{M}= J_{0}(\lambda ) .
	\label{eq:M}
\end{gather}
The MQO correspond to the rapidly oscillating factors $\sin\left(\overline{\alpha} \right)$ or $\cos \left(\overline{\alpha} \right)$ in Eqs. (\ref{eq:M0})-(\ref{eq:M}), while the factors $R_{M}$ and $R_{z\sigma }$ describe the beats of MQO amplitudes and are the subject of our study.

\section{Analysis and Results}

Let us first consider the limit of $\lambda\gg 1$, corresponding to a rather large electron interlayer transfer integral $4\pi t_z\gg \hbar\omega_c$ or weak magnetic field. Then one can use the large-argument asymptotic expansions of the Bessel functions in Eqs. (\ref{eq:szzqo}) and (\ref{eq:M}). As a result, the beating factors of the MQO amplitudes in Eqs. (\ref{eq:szzqo}) and (\ref{eq:M}) simplify to
\begin{gather}
R_{M}= \sqrt{\frac{2}{\pi }} \frac{\cos\left(\lambda-\frac{\pi}{4}\right) }{\sqrt{\lambda }},
\label{eq:M-cos} \\
 R_{z\sigma }= \sqrt{\frac{2}{\pi }} \frac{ \cos \left(\lambda-\frac{\pi }{4}\right)}{\sqrt{\lambda}}+\frac{(16 \gamma_0+15) \cos \left(\lambda+\frac{\pi }{4}\right)}{4 \sqrt{2 \pi }\lambda^{3/2}}. \label{eq:szzqo-1}
\end{gather}
At $\lambda \gg 2(\gamma_0+1)$ the second term in Eq. (\ref{eq:szzqo-1}) is negligibly small, and we get the same beating factors of the MQO of magnetization and interlayer conductivity,
\begin{gather}
	R_{z\sigma }\approx \frac{\sqrt{\frac{2}{\pi }} \cos \left(\lambda-\frac{\pi }{4}\right)}{\sqrt{\lambda}} = R_{M}.
	\label{eq:szzqo-2}
\end{gather}
Hence, in this limit the phase shift of beats is absent, $\phi_b\approx 0$. 

In the opposite limit of strong MQO damping by disorder, 
$\gamma_0\gg\lambda /2$, the first term in Eq. (\ref{eq:szzqo-1}) can be omitted, and one obtains
\begin{gather}
R_{z\sigma }\approx \frac{2^{3/2} \gamma_0 \cos \left(\lambda+\frac{\pi }{4}\right)}{ \sqrt{\pi }\lambda^{3/2}}.
	\label{eq:SzzQO-cos-large-gamma}
\end{gather}
Comparing Eqs. (\ref{eq:M-cos}) and (\ref{eq:SzzQO-cos-large-gamma}) we see that at $\gamma_0\gg\lambda /2$ the MQO beats of interlayer conductivity are shifted from those of magnetization by the phase $\phi_b=\pi /2$. 
As follows from Eq. (\ref{eq:M-cos}), the beat nodes of magnetization oscillations $\widetilde{M}$ in Eq. (\ref{eq:M}) are located at $\lambda\approx-\pi/4+\pi n$, and from Eq. (\ref{eq:SzzQO-cos-large-gamma}) it follows that the beat nodes of the interlayer conductivity oscillations $\sigma_{zz}^{QO}$ given by Eq. (\ref{eq:szzqo}) are shifted by $\pi /2$ and located at $\lambda\approx\pi/4+\pi n$, where $n\geq 1$ is an integer number. These $\lambda$ give the same position of the beat nodes as the expressions (\ref{eq:Bnote-dHvA}) and (\ref{eq:Bnote-SdH}) for $n\geq 1$ and $\Delta F=\lambda |B_z|/\pi$, corresponding to the fit of experimental data in Ref. \cite{Schiller2000Jul}. Hence, Eqs. (\ref{eq:M-cos}) and (\ref{eq:SzzQO-cos-large-gamma}) describe well the experimental observations in Ref. \cite{Schiller2000Jul}. However, Eq. (\ref{eq:SzzQO-cos-large-gamma}) assumes $\gamma_0\gg\lambda /2$, while $\gamma_0\sim \lambda $ in the experiment in Ref. \cite{Schiller2000Jul}. 
The Dingle temperature extracted from the experimental data in Ref. \cite{Schiller2000Jul} is $T_{D}\approx 0.5$ K. The beat frequency in the organic metal (BEDT-TTF)$_4$[Ni(dto)$_2$] studied in Ref. \cite{Schiller2000Jul} is $\Delta F \approx 4.5$ T, which corresponds to the interlayer transfer integral $t_z=0.06$ meV $\approx 0.7$ K. Hence, the ratio $\gamma_0/\lambda = 2\pi^2 T_{D}/(4\pi t_z) = \pi T_{D}/(2 t_z)\approx 1.12$ in Ref. \cite{Schiller2000Jul}. 
Hence, to improve the quantitative description of the observed phase shift of MQO beats we need to consider the case $ \lambda \sim \gamma_0\sim 1$ more accurately.  

\begin{figure}[bht]
		\centering
		\includegraphics[width=\textwidth]{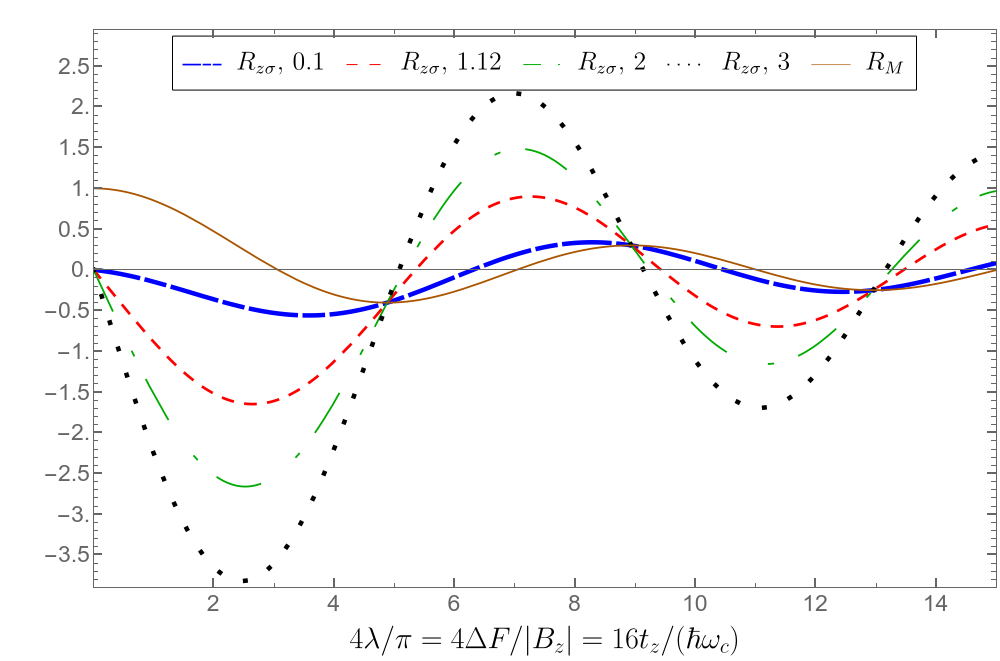}
		\caption{The beating factors $R_{z\sigma}\equiv \left[J_{0}\left(\lambda\right)-\left(1+\gamma_{0}(\lambda)\right)J_{1}\left(\lambda\right)/\lambda\right]$ and $R_M \equiv  J_{0}(\lambda)$, describing the MQO amplitudes of interlayer conductivity $\sigma_{zz}^{QO}$ and magnetization $\widetilde{M}$ (brown solid curve), as a function of the inverse magnetic field $1/|B_z|$. Here $\Delta F=4t_zB_z/(\hbar\omega_c)$ is the beat frequency, and the ratios $\gamma_0/\lambda = 0.1,\,1.12,\,2,\,3 $ correspond to the blue thick long-dashed, red short-dashed, green long-dashed and black dotted curves for $R_{z\sigma}$.}
	\label{fig:1}
\end{figure}

The phase shift $\phi_b$ of MQO beats becomes even stronger than that given by Eqs. (\ref{eq:M-cos}) and (\ref{eq:SzzQO-cos-large-gamma}) if one applies Eqs. (\ref{eq:szzqo}) and (\ref{eq:M}) without the asymptotic expansions of the Bessel functions at large argument, i.e. without the assumption $\lambda \gg 1$, which is more relevant to the experiment in Ref. \cite{Schiller2000Jul}. In Fig. \ref{fig:1}, using Eqs. (\ref{eq:szzqo}) and (\ref{eq:M}) at $\gamma_0(\lambda)=\{0.1,\,1.12,\,2,\,3\}\lambda$, we plot the beating factors $R_{M}$ and $R_{z\sigma }$, describing the MQO amplitudes  of magnetization and conductivity without the Dingle and temperature damping factors. The theoretical curves in Fig. \ref{fig:1} correspond to four values $\gamma_0/\lambda = \{0.1,\,1.12,\,2,\,3\}$ to describe the experimental data from Refs. \cite{Grigoriev2002Jan} and \cite{Schiller2000Jul} where $\gamma_0/\lambda \approx 0.1$ and $\gamma_0/\lambda \approx 1.12$ correspondingly. In Fig. \ref{fig:1} we take a special scale of the abscissa axis so that the zeros of both curves, called the beat nodes, fall into integer values. From Fig. \ref{fig:1} we see that the beat nodes for interlayer conductivity oscillations $\sigma^{QO}_{zz}$ are shifted relative to the magnetization $\widetilde{M}$ beat nodes by a quarter of the period $\lambda\approx\pi\Delta F/|B_z|$.\footnote{In \cite{Mogilyuk2018Jul} we used different definition of $\Delta F\approx2 t_zB_z/(\hbar\omega_c)=\lambda B_z/\pi$.} The beat nodes of magnetization oscillations $\widetilde{M}$ given by Eq. (\ref{eq:M}) in Fig. \ref{fig:1} are located at $4\lambda/\pi \approx\{3,\,7,\,11,\,15\}$. This coincides with the prediction of simplified limiting-case formula (\ref{eq:M-cos}) and agrees perfectly with the experimental data from Ref. \cite{Schiller2000Jul}, fitted by Eq. (\ref{eq:Bnote-dHvA}). 
At $\gamma_0(\lambda) = 0.1\lambda$, the positions of beat nodes of the amplitude of conductivity oscillations are close to the position of the beat nodes of the amplitude of magnetization oscillations. The beat nodes of interlayer conductivity oscillations $\sigma_{zz}^{QO}$ given by Eq. (\ref{eq:szzqo}) and plotted in Fig. \ref{fig:1} at $\gamma_0(\lambda)=\{1.12,\,2,\,3\}\lambda$ are located at $4\lambda/\pi\approx\{0,\,5,\,9,\,13\}$.
For $n\geq 1$ they coincide with Eq. (\ref{eq:Bnote-SdH}), which fits the experimental data from Ref. \cite{Schiller2000Jul}.

\begin{figure}[tbh]
		\centering
		\includegraphics[width=\textwidth]{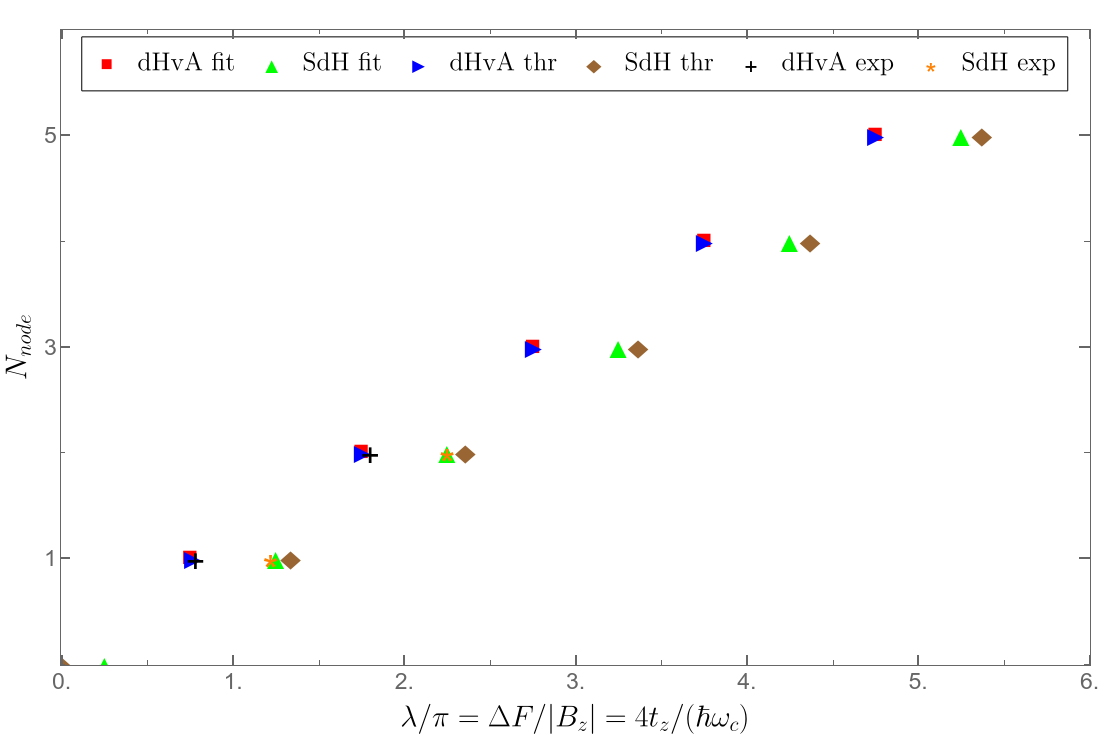}
	\caption{The dependence of the beat-node number $N_{node}$ of the MQO of $\sigma_{zz}^{QO}$ and $\widetilde{M}$ on the inverse magnetic field $1/|B_z|$ at $\gamma_0 =1.12\lambda$.
	Points <<dHvA fit>> and <<SdH fit>> refer to the fit expressions (\ref{eq:Bnote-dHvA}) and (\ref{eq:Bnote-SdH}) from Ref.
	\cite{Schiller2000Jul}. 
	Points <<dHvA exp>> and <<SdH exp>> refer to the experimental data from Figs. 2 and 3 of Ref. \cite{Grigoriev2002Jan}.
	Points <<dHvA thr>> are found from the zeros of expression (\ref{eq:M-cos}), while points <<SdH thr>> are found numerically from the solution of equation $R_{z\sigma }= J_{0}\left(\lambda\right)-\left(1+\gamma_{0}(\lambda)\right)J_{1}\left(\lambda\right)/\lambda=0$.}
	\label{fig:2}
\end{figure}

Eqs. (\ref{eq:szzqo}) and (\ref{eq:M}) also perfectly agree with the experimental data from Ref. \cite{Grigoriev2002Jan}. In Fig. \ref{fig:2} we plot the measured positions of magnetization and conductivity beat nodes from the experiments in Refs. \cite{Grigoriev2002Jan} and \cite{Schiller2000Jul} together with the theoretical predictions from formulas (\ref{eq:szzqo}) and (\ref{eq:M}). The measured positions of the beat nodes in Ref. \cite{Schiller2000Jul} at $n\geq 1$ are very well described by Eqs. (\ref{eq:Bnote-dHvA}) and (\ref{eq:Bnote-SdH}) and, therefore, denoted by ''dHvA fit'' and ''SdH fit'' in Fig. \ref{fig:2}. The points ''dHvA exp'' and ''SdH exp'' in Fig. \ref{fig:2} are taken from the experimental data in Figs. 2 or 3 of \cite{Grigoriev2002Jan}. The prediction of Eqs. (\ref{eq:szzqo}) and (\ref{eq:M}) are denoted by ''SdH thr'' and ''dHvA thr'' in Fig. \ref{fig:2}. From Fig. \ref{fig:2} we see a perfect agreement between the above theory and the experimental data from Refs. \cite{Grigoriev2002Jan} and \cite{Schiller2000Jul} for the beat-node positions with number $n\geq 1$. These beat-node positions for the MQO of magnetization and interlayer conductivity are fitted well by two straight lines parallel to each other.
Hence, we conclude that Eqs. (\ref{eq:M}) and (\ref{eq:szzqo}) describe very well the beating nodes with $n\geq 1$.

The position $\lambda=0$ of the zeroth beating node given by Eq. (\ref{eq:szzqo}) holds at any ratio $\gamma_0 /\lambda$, as illustrated in Fig. \ref{fig:1} and can be shown analytically. Indeed, in a very high magnetic field $\lambda\to 0$, the ratio $\gamma_0 /\lambda = \hbar/(4\tau t_z) \to const$, and the beating nodes are given by the equation
\begin{equation}
	R_{z\sigma }= J_{0}\left(\lambda\right)-\frac{2}{\lambda}%
	\left(1+\gamma_{0}\right)J_{1}\left(\lambda\right) \approx - \frac{\gamma_0}{\lambda }\lambda  + \frac{\left(\gamma_{0}-1\right)}{8}\lambda^2 =0.  \label{Rzs}
\end{equation}
$\lambda =0$ is the root of this equation at any ratio $\gamma_0 /\lambda$. The second root $ \lambda =\sqrt{8\gamma_0 /\left(\gamma_{0}-1\right)}$ of Eq. (\ref{Rzs}) is physically irrelevant as it gives a too large value 
\begin{equation}
	\lambda =\frac{1+\sqrt{1+32\gamma _{0}^{2}/\lambda ^{2}}}{2\gamma
		_{0}/\lambda }>2\sqrt{2}
\end{equation}
much beyond the limit $\lambda \ll 1$ where the expansion (\ref{Rzs}) holds.

The position of the very first beat node at $n=0$ differs in Eqs. (\ref{eq:Bnote-SdH}) and (\ref{eq:szzqo}).  
According to Eq. (\ref{eq:Bnote-SdH}), the beating node $N_{node}=0$ of $\sigma^{QO}_{zz}$ oscillations should be at $4\lambda/\pi=1$, which does not coincide with the position $B=\infty$ or $\lambda =0$ of the zeroth beating node given by Eq. (\ref{eq:szzqo}) and shown in Fig. \ref{fig:1}. As we just have shown, the origin of this difference is not a ''wrong'' choice of the parameter $\gamma_0 /\lambda$, entering Eq. (\ref{eq:szzqo}). 

In fact, the position $\lambda =\pi /4$ of the zeroth beating node $n=0$, predicted by the ''fit'' Eq. (\ref{eq:Bnote-SdH}), has not been observed experimentally. The beat frequency in the organic metal (BEDT-TTF)$_4$[Ni(dto)$_2$] studied in Ref. \cite{Schiller2000Jul} is $\Delta F \approx 4.5$ T. According to  Eq. (\ref{eq:Bnote-SdH}) the zeroth beating node is at $B=4\Delta F \approx 18$ T. In Ref. \cite{Schiller2000Jul} the experimental data on the SdH oscillations in (BEDT-TTF)$_4$[Ni(dto)$_2$] are shown up to $B=6$ T only in Fig. 6 of Ref. \cite{Schiller2000Jul}, while the data up to $B=28$ T are given only for dHvA oscillations in Fig. 3 of Ref. \cite{Schiller2000Jul}. In Fig. 6.1 of Ref. \cite{Balthes2004} the SdH data up to $B=28$ T are shown for the same organic metal (BEDT-TTF)$_4$[Ni(dto)$_2$], and no beating nodes are observed in the magnetic-field interval $4<B<28$ T. 
A similar difference in the beat-node positions of de Haas - van Alphen and Shubnikov oscillations in
$\kappa-\textrm{(BEDT-TTF)}_2\textrm{Cu[N(CN)}_2]\textrm{Br}$ \cite{Weiss1999Dec} was also detected by the non-observation of the SdH beat node in the magnetic-field interval $16<B<28$ T. Hence, Eq. (\ref{eq:szzqo}) describes very well the available experimental data on the MQO beats of interlayer conductivity.

\begin{figure}[H]
		\centering
		\includegraphics[width=\textwidth]{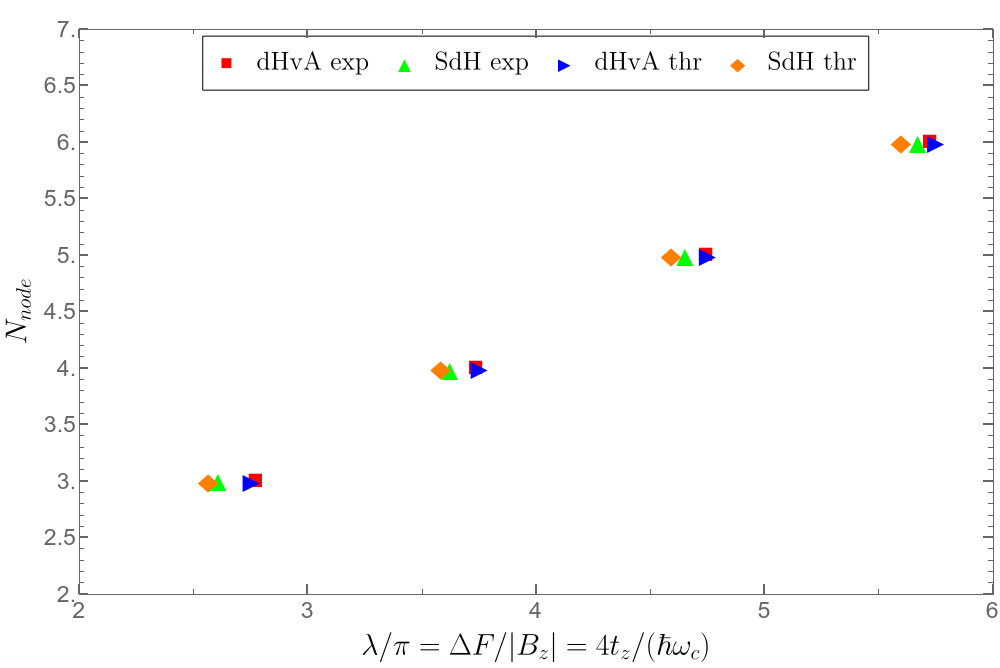}
	\caption{The dependence of the number $N_{node}$ of the beat node of magnetic quantum oscillations of $\sigma_{zz}^{QO}$ and $\widetilde{M}$ on the inverse magnetic field $1/|B_z|$ at $\gamma_0(\lambda)=0.1\lambda$. 
	The points <<dHvA exp>> and <<SdH exp>> refer to the experimental points taken from the Fig. 3 of \cite{Grigoriev2002Jan}.
	The points <<dHvA thr>> are found from the zeros of expression (\ref{eq:M-cos}), while the points <<SdH thr>> are found numerically from the solution of equation $R_{z\sigma }= J_{0}\left(\lambda\right)-\left(1+\gamma_{0}(\lambda)\right)J_{1}\left(\lambda\right)/\lambda=0$. Here $\Delta F=4t_zB_z/(\hbar\omega_c)$.}
	\label{fig:3}
\end{figure}

In Ref. \cite{Grigoriev2002Jan} the Dingle temperature extracted from the experimental data is $T_D=0.8$ K, but the Dingle temperature from the scattering by short-range disorder is $T_D^*=0.15$ K (see Fig. 4 of Ref. \cite{Kartsovnik2002Aug}). The beat frequency $\Delta F \approx 40.9$ T in the organic metal $\beta $-(BEDT-TTF)$_2$IBr$_2$ studied in Ref. \cite{Grigoriev2002Jan} is much larger than in (BEDT-TTF)$_4$[Ni(dto)$_2$] studied in Ref. \cite{Schiller2000Jul}. It corresponds to the interlayer transfer integral $t_z=1.15$ meV $\approx 13.3$ K and $\gamma_0/\lambda \approx 0.1$.
For $\gamma_0\ll\lambda/2^{3/2}$ the expression for $R_{z\sigma }\left(\lambda\right) $ in Eq. (\ref{eq:szzqo-1}) is approximately equal to that in Eq. (\ref{eq:szzqo-2}).
Since formulas (\ref{eq:M-cos}) and (\ref{eq:szzqo-2}) are similar, this means that the beat nodes for magnetization $\widetilde{M}$ and conductivity oscillations $\sigma^{QO}_{zz}$ in this limit are close to each other. In Fig. \ref{fig:3} we show that the beat nodes found at $\gamma_0=0.1 \lambda$ fit well with the experimental data on the position of beat nodes from Fig. 3 of Ref. \cite{Grigoriev2002Jan}. 
The higher is the number $N_{node}$ of the MQO beat nodes of magnetization and interlayer conductivity, the shorter is the distance between them along the abscissa axis in the units of $\Delta F/|B_z|$.

\section{Discussion}

The above analytical formulas describe the interlayer conductivity $\sigma _{zz}$. They can also be applied to describe the first MQO harmonic of interlayer magnetoresistance $\rho_{zz} =1/\sigma _{zz}$, if the MQO are not very strong so that the non-linear effects in MQO amplitude are not important \cite{Grigoriev2023Hall}. If only the first MQO harmonic is kept and the higher-order terms in the MQO amplitude are negligible, one can safely apply the relation $\rho_{zz} =\sigma^{-1} _{zz}$ to describe the interlayer magnetoresistance \cite{Grigoriev2023Hall}. However, for strong MQO their averaging over temperature and long-range disorder differs \cite{Grigoriev2012Oct,Grigoriev2023Hall} for $\rho_{zz}$ and $\sigma _{zz}$. 

\begin{figure}[H]
		\centering
		\includegraphics[width=\textwidth]{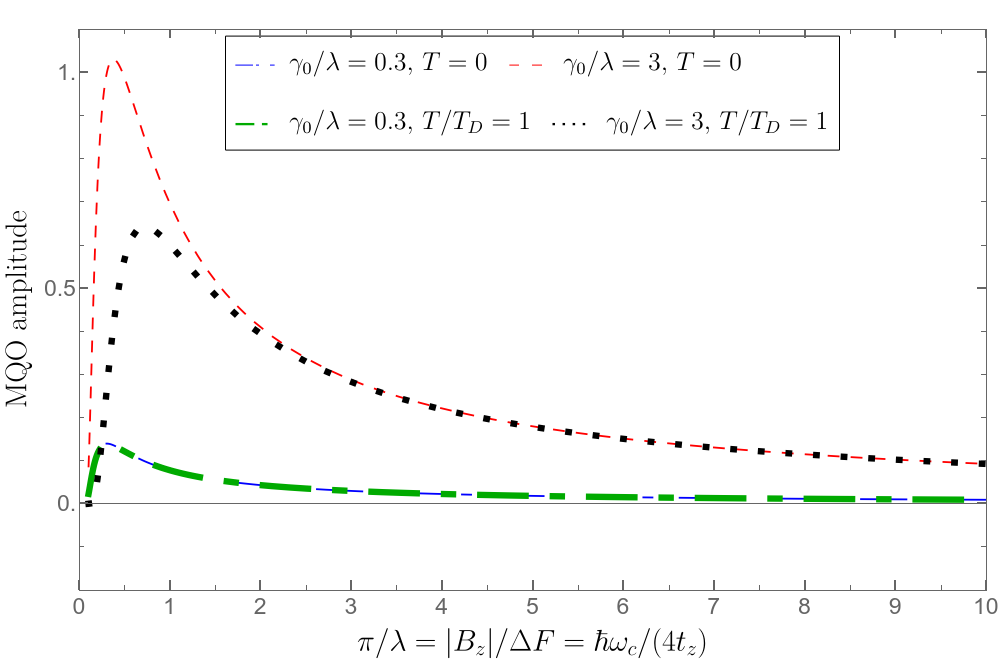}
	\caption{Dependence of the value $(-1)R_{D0}R_{T} R_{z\sigma }$ on the magnetic field for $\gamma_{0}=\{0.3,\,3\}\lambda$ for temperature $T=\{0,\,1\}T_D$, where Dingle temperature is equal to $T_D\equiv\hbar/(2\pi k_B \tau)$.}
	\label{fig:4}
\end{figure}

Above we have shown that the developed theory and Eq. (\ref{eq:szzqo}) describe very  well the available experimental data on the MQO beats of interlayer conductivity $\sigma _{zz}$. They also predict some new interesting and observable features. In particular, Eq. (\ref{eq:szzqo}) predicts a strong decrease of the MQO beating factor $R_{z\sigma}$ of interlayer conductivity with the increase of magnetic field, corresponding to $\lambda \to 0$. This decrease of $R_{z\sigma}$ is partially compensated by the Dingle factor $R_{D0}$ and by the temperature damping factor $R_T$. However, the product $R_{D0}R_{T} R_{z\sigma }$, describing the MQO amplitude of interlayer conductivity, may still be non-monotonic in high field, in contrast to the amplitude of magnetization MQO described by Eq. (\ref{eq:M}), which always monotonically increases in a very high field at $\hbar\omega_c\gg t_z$. This nonmonotonic magnetic-field dependence of the MQO amplitude of interlayer conductivity in a high field is rather robust to the variation of material parameters, e.g, to variations of the ratio $\gamma_{0}/\lambda = \pi k_B T_D/(2t_z)=\hbar /(4t_z\tau )$. 
In Fig. \ref{fig:4} we show this non-monotonic dependence of the $R_{zz}$ MQO amplitude $(-1)R_{D0}R_{T} R_{z\sigma}$ on $\pi/\lambda=|B_z|/\Delta F$ for two different ratios $\gamma_0/\lambda $ and temperatures $T$. At the small ratio $\gamma_0/\lambda < 0.1$ the increase of temperature up to the Dingle temperature $T_D$ has almost no effect on the graph. 
The predicted decrease of MQO amplitude of the interlayer conductivity in a high field is somewhat counterintuitive, being opposite to what is observed in magnetization, but it can be easily tested experimentally in various Q2D layered compounds, such as organic metals, van der Waals crystals, artificial heterostructures, etc. We suppose that the proposed decrease of MQO amplitude in a high field was actually observed in many experiments but mistaken for the non-existing beat node at a high magnetic field beyond the available range. For example, as we mentioned in the previous section, in Refs. \cite{Weiss1999Dec,Schiller2000Jul,Balthes2004} the beating node $n=0$ of the SdH effect was not observed but assumed from some decrease of MQO amplitude in a very high field. This decrease may be considered as the experimental indication of the predicted high-field non-monotonic behavior of the MQO amplitude of interlayer conductivity, but further experimental studies of this interesting effect are required to compare with our theoretical predictions and to determine the range of parameters where it is observed.

Another amazing prediction of our theoretical analysis is that all the curves in Fig. \ref{fig:1}, describing the MQO amplitudes, periodically cross at the same points. It can also be checked experimentally. However, this experimental test is less obvious and convenient than the above prediction of the monotonic magnetic-field dependence of $R_{zz}$ MQO amplitude, because the corresponding experimental data for $M(B_z)$ and $R(B_z)$ magnetic oscillations must be properly normalized for this comparison.

Now we consider the MQO phase of interlayer conductivity. For the Q2D electron spectrum (\ref{e3Dg}) in a magnetic field one can easily calculate \cite{Champel2001Jan,Grigoriev2001Jun} the one-particle DoS $\rho $. If we neglect higher harmonics, the DoS oscillations are given by \cite{Champel2001Jan,Grigoriev2001Jun}
\begin{equation}
	\widetilde {\rho} \approx \rho _{0}2R_{D}R_{\rho}\cos \left( \bar{\alpha}\right), \ \ R_{\rho}=-J_{0}(\lambda ),
	\label{eq:rho}
\end{equation}%
where $\rho_{0}= m_{\ast}/(2\pi\hbar^{2}d)$ is its nonoscillating part per one spin component. The phase of DoS oscillations is strictly tied to that of magnetization given by Eq. (\ref{eq:M}), because the beating factors $R_{M}=-R_{\rho}$. In a weak magnetic field, when $\lambda \gg 1$ corresponding to the 3D limit, the second term in the beating factor $R_{z\sigma}$ of interlayer conductivity given by Eq. \ref{eq:szzqo} is much smaller than its first term. Then we have $R_{z\sigma}\approx R_{M}=J_{0}(\lambda ) =-R_{z\rho}$, and the MQO phases of the interlayer conductivity $\sigma _{zz}$ and of the DoS $\rho$ are strictly opposite, as predicted by the 3D theory of SdH effect \cite{Abrikosov1988Fundamentals,Ziman1972Principles}. In the opposite case $\lambda < \pi $, corresponding to the 3D $\to $2D crossover or nearly 2D limit, the SdH beating factor $R_{z\sigma}<0$ is negative, as follows from Eq. \ref{eq:szzqo} and illustrated in Fig. \ref{fig:4}. Since $R_{\rho}=-R_{M}<0$ is also negative at  $\lambda < 1$, at the 3D $\to $2D crossover $B\gtrsim\Delta F$ or in the almost 2D limit $B\gg \Delta F$ the quantum oscillations of the electronic DoS and of the interlayer conductivity have the same phase. We see that our theory describes the phase inversion of the SdH oscillations during the 3D $\to $2D crossover. 

The above analysis is generic to layered Q2D metals and can be applied to a very large class of layered compounds, which are actively studied by MQO till nowadays \cite{Oberbauer2023,Xu2020,Helm2021,Luo2024,Danica2023,Rehfuss2024,Sakai2023,Zeng2022,Zhang2022}. 
To compare our predictions with experiment it is sufficient to measure the MQO of interlayer conductivity, but a full comparison, including the phase shift, is possible if the MQO of another quantity are also measured. This may be any thermodynamic quantity, such as magnetization or heat capacity, sound attenuation, or another transport quantity, such as intralayer magnetoresistance or thermal conductivity.

\section{Conclusions}

Above we developed the quantitative theoretical description of the beating phase of magnetic quantum oscillations of interlayer magnetoresistance and magnetization in quasi-2D metals, which are traditionally used to extract useful information about the electronic properties. This theory agrees very well with the available experimental data and explains the long-standing puzzle of the observed too large phase shift of beats between the MQO of thermodynamic and transport electronic properties. In addition, our theory also makes several new predictions which can be tested experimentally. The most unexpected prediction is the non-monotonic field dependence of the MQO amplitude of interlayer magnetoresistance in a high magnetic field $B_z\sim \Delta F$, where $\Delta F$ is the beat frequency. Naively one may expect a monotonic growth of MQO amplitude of interlayer conductivity in a strong magnetic field $|B_z| > \Delta F$, similar to magnetization, but we predict its decrease at $|B_z|>\Delta F$ in the regime of 3D $\to $ 2D crossover, as illustrated in Fig. \ref{fig:4}. Our analytical formulas also explain and describe the phase inversion of the SdH oscillations during the 3D $\to $2D crossover. 

\vspace{6pt}

\authorcontributions{Conceptualization, P.D.G.; methodology,  P.D.G. and T.I.M.; validation, I.V., V.D.K. and T.I.M.; formal analysis, T.I.M.; investigation, P.D.G. and T.I.M.; writing---original draft preparation, T.I.M.; writing---review and editing, P.D.G.; supervision, P.D.G.; funding acquisition, I.Y.P. and P.D.G. All authors have read and agreed to the published version of the manuscript.}

\funding{This research was funded by the Russian Science Foundation grant \#22-42-09018. V.D.K. acknowledges the Foundation for the Advancement of Theoretical Physics and Mathematics ”Basis” for grant \# 22-1-1-24-1 and NUST "MISIS" grant No. K2-2022-025.}

\dataavailability{Dataset available on request from the authors.} 



\conflictsofinterest{The authors declare no conflict of interest.} 



\abbreviations{Abbreviations}{
The following abbreviations are used in this manuscript:\\

\noindent 
\begin{tabular}{@{}ll}
MQO & Magnetic quantum oscillations\\
Q2D & Quasi-two-dimensional \\
FS & Fermi surface \\
LK & Lifshitz-Kosevich \\
dHvA & de Haas - van Alphen \\
SdH & Shubnikov - de Haas
\end{tabular}
}

\begin{adjustwidth}{-\extralength}{0cm}

\reftitle{References}


\bibliography{Papers}

\PublishersNote{}
\end{adjustwidth}
\end{document}